# SENSAAS (SENsitive Surface As A Shape): utilizing open-source algorithms for 3D point cloud alignment of molecules

*Dominique Douguet[§]\* and Frédéric Payan[±]\**

[§]Université Côte d'Azur, Inserm, CNRS, IPMC, 660 route des lucioles 06560 Valbonne, France

[±]Université Côte d'Azur, CNRS, I3S, Les Algorithmes - Euclide B, 2000 route des lucioles 06900 Sophia Antipolis, France

**ABSTRACT** Open-source 3D data processing libraries originally developed for computer vision and pattern recognition are used to align and compare molecular shapes and sub-shapes. Here, a shape is represented by a set of points distributed on the van der Waals surface of molecules. Each point is colored by its closest atom, which itself belongs to a user defined class. The strength of this representation is that it allows for comparisons of point clouds of different kind of chemical entities: small molecules, peptides, proteins or cavities (the negative image of the protein surface). The SENSAAS (SENsitive Surface As A Shape) workflow we developed for aligning two molecules follows three major steps. First, it begins by generating surfaces and derived colored point clouds. Secondly, a *Global registration* method is executed to generate initial superimpositions. For two given point clouds, it consists in i) down-sampling them, ii) computing a local descriptor named FPFH for each remaining point to get a comprehensive description of local shape geometry, iii) finding the best matching between the features of each point cloud by using a RANSAC-based method. Thirdly, these initial superimpositions are refined by using the *Colored point cloud registration* method on the same down-sampled, but colored, point clouds, in order to take into account for the physico-chemical properties when optimizing the superimpositions. In this study, parameters were optimized to calibrate Open3D registration methods to molecular shapes. SENSAAS provides a score and a pairwise alignment. It evaluates the molecular similarity by using three Tversky coefficients (ranging from 0 to 1). To show its utility, we investigated the alignment of several test cases ranging from similar to dissimilar molecules, conformers, bioisosteric chemical groups, substructures and fragments.

**INTRODUCTION**

Representing a molecule by using a 3D point cloud is not a new concept (1-5). A point cloud has this advantage of being an intuitive description suited to the visualization and to build a mental model. This representation also allows coloring points to highlight local shape properties and distributions, like a 3D pharmacophore that would also consider the three-dimensional form. Although they are simple to generate, point clouds may be difficult to align. *Point set alignment*, also known as *registration* or *matching* is a fundamental problem in many domains such as robotics, pattern recognition, computer vision or data reconstruction (6). Alignment is often considered as a hard optimization problem because perfect point-to-point correspondences rarely exist: the point clouds might partially overlap only, or the underlying objects may have particular local geometrical features. A large number of solutions have been proposed for matching point clouds since decades (7,8). In our study, we investigated if newly registration methods may contribute to evaluate the molecular similarity of molecules when they are represented by colored point-based surfaces. Point-based surface methods are a particular class of shape similarity methods that are still in infancy (3,4).

**METHODS**

We mainly investigated registration methods developed in the Open3D library (9), freely available at the web site [www.open3d.org](www.open3d.org). Based on these methods, we developed SENSAAS. Considering two molecules, our alignment method follows four main steps described hereafter: i) generating point clouds describing the molecular surface; ii) assigning a color to each point according to several user-defined classes; iii) applying a geometry-aware registration for aligning the two point clouds globally; iv) applying color and geometry-aware local registration for refining the result of the first step.

SENSAAS is a rigid-body superimposition method whose results depend on the 3D conformation of the molecules. SDF or PDB format files are accepted as inputs in our method. For small molecules or peptides, we highly recommend generating an ensemble of conformers in order to retrieve different shapes and 3D point clouds. Indeed, the more flexible the molecules are, the more important an ensemble of conformers is. The e-LEA3D web server (10) offers a service to draw and build ensemble of conformers by using the RDKit Open-Source Cheminformatics Software (http://rdkit.org).

**I - Creation of 3D point clouds**

**1) Generating Molecular Surfaces**

The nsc program developed by F. Eisenhaber *et al.* is used to calculate the van der Waals (vdW) surface of an input molecule (11), by using van der Waals radii taken from A. Bondi (12). It results into a cloud of points described by their 3D coordinates, and uniformly distributed on the underlying vdW surface. Each point is then labeled with the element name of its closest atom and saved in a PDB format file.

**2) Hydrogen atoms and the molecular shape**

We highly recommend to protonate input molecules. Hydrogen atoms are numerous in chemical entities. They significantly affect the geometry of the molecular shape and also contribute to the labeling of points distributed on the vdW surface. Also, hydrogen atoms often hide skeleton atoms like carbons. For example, in **Figure 1**, we show the 3D graph structure and the corresponding point cloud of the compound indoxam with (**Fig. 1** a, b, c, d and e) and without (**Fig. 1** f and g) hydrogens (hydrogen atoms are colored with white sticks and points). We can observe that the shape of the molecule differs significantly between the two states as well as the distribution of colors on the surface (**Fig. 1** b and g). In the case of the protonated indoxam, aromatic carbons contribute to the surface with a typical pattern formed by two parallel green patches (**Fig. 1** b). Such features are easily identified by the matching algorithm.

Of note, as hydrogens linked to nitrogens and oxygens are usually polar atoms, we categorize them into the same polar group of atoms (see **I-3**). It means that such hydrogen atoms will possess the same color as the attached nitrogen or oxygen atom (**Fig. 1** c compared to b).

Tautomers are distinguished by the migration of a hydrogen atom on the structure and ionized molecules usually differ by the addition or removal of one hydrogen atom (for example, carboxylic acid / carboxylate forms of a molecule). Such changes have limited effects on the alignment results with our method because the difference in the number of colored points is small (**Fig. 1** c and e). However, these different molecular states may lead to different conformers and, thus, may lead to different colored point clouds.

**3) Color assignment of points in the cloud**

In this study, we categorize chemical elements into four color classes, according to their physico-chemical properties. These color classes will help to refine the alignment of the input molecules during the second registration step and, also, to set several scores:

- **class 1 [H, Cl, Br and I].** This first class includes non polar hydrogens and halogens, excepting fluorine. Hydrogens and halogens are molecule endings: these elements significantly contribute to the surface geometry and coloration. This class is usually the most populated. It outlines the global geometry and shows the distribution of apolar surface areas. This class is colored in white in our study.

- **class 2 [N, O, S, H(NH/OH), F]**. This second class includes polar atoms able to be involved in hydrogen bonds. Class 2 colored patches display polar surface areas on the molecular surface. We also categorize polar hydrogens and fluorine elements in this class (13). This class is colored in red in our study.

- **class 3 [C, P, B]**. This third class includes skeleton elements. Sometimes they contribute to the surface. For example, a sp3 carbon does not contribute much to the surface because linked hydrogen atoms masked it. On the contrary, aromatic carbons may contribute to the surface as shown in Figure 1 (**Fig. 1** b). As this class is colored in green in our study, such aromatics display a typical pattern formed by two parallel green patches.

- **class 4 [other atoms]**. This fourth class includes all other elements not listed in the other classes. This class is empty for most small organic molecules in medicinal chemistry. This class is colored in blue in our study.

These four classes recapitulate the major physico-chemical features that usually describe a pharmacophore (defined as the common features of known bioactive compounds): aromatic center, lipophilic center, hydrogen-bond donors and hydrogen-bond acceptors. In our study, these pharmacophoric features are distributed on the vdW surface instead of using the 3D graph coordinates. For example, such classification allows our method to display biosisosterism between a tetrazole and a carboxylate function (see RESULTS).

Of note, defining too many classes, for example by choosing one color per atom, may reduce the efficiency of the *Colored point cloud* registration method (step 4), because that will generate many small patches all over the surface.

**II - Alignment**

Open3D is an open-source library that supports the development of software dealing with 3D data (9) (https://arxiv.org/abs/1801.09847). We use successively two complementary registration methods available in the library: a *Global geometry-aware registration* method (named GLOBAL, from now on) that produces an initial alignment, and the *Colored point cloud registration* method (named COLOR, from now on) that further refines the superimposition. 3D registration methods are calculated using PCD (Point Cloud Data) format files as inputs. We use BioPandas python library to read PDB input point clouds and then convert them into PCD format files (http://pointclouds.org/documentation/tutorials/pcd_file_format.php). A PCD aggregates 3D coordinates of points and an associated color (RGB), depending on the class of the point (color of the closest atom class). During the superimposition of two point clouds, we set the first input molecule as *Target* and the second input molecule as *Source*. The *Source* point cloud is rotated and translated to optimally match the *Target* point cloud.

**a) GLOBAL**

GLOBAL is feature-based registration. The base concept is to compute a multi-dimensional descriptor for each point of the clouds, to describe the local geometry of the inherent surfaces. The registration thus consists in finding the best matching between the descriptors of each point cloud. This matching provides a matrix transformation that describes the rotation and the translation that must be applied on the *Source* to be aligned on the *Target*.

In this method, the descriptors are based on histograms, the Fast Point Features Histograms, aka FPFH (14). These histograms provide pose invariant local properties of the inherent surface geometry. In a nutshell, according the shape of its histogram, we can know if the surface around a point looks like a sphere, an angle, a plane…

Once *Source* and *Target* point clouds described separately by two sets of FPFH, RANSAC is applied (15). RANSAC has been developed during the eighties to estimate parameters of a mathematical model from a set of observed data iteratively. In our context, RANSAC is used to compute the elements of the matrix transformation that permits to "go from the source to the target" with only a translation and a rotation. Once this matrix applied, the two point clouds are registred, and finally we have the "best" alignment for the associated molecules. For more details, we suggest the readers to study the work of Choi *et al.* (16).

**b) COLOR**

COLOR is a local registration method that considers geometry but also color of the point clouds. This method is particularly relevant for matching points sets acquired with RGB-D sensors for instance. Such sensors are used to capture real life scenes: they are able to provide not only a 3D point cloud that describes the surface geometry, but also to capture the color to each point. Given two colored point sets, the COLOR method finally tries to find the matrix transformation that minimizes iteratively and jointly a functional combining the 3D coordinates and the RGB information of each point.

As explained before, for aligning molecular surfaces, we consider the influence of the atoms on the surfaces by considering user-defined classes. Each class being defined by a color associated to each point, we can benefit from the COLOR approach to improve the initial alignment obtained with the GLOBAL approach. For more details, we suggest the readers to study the work of Park *et al.* (17).

**III - Evaluation scores**

After performing the alignment of the *Source* and the *Target* as explained previously, SENSAAS evaluates its efficiency with *fitness scores*. The Open3D library includes a function called *evaluate registration* that calculates the amount of matching points between *Source* and *Target*. As it evaluates a geometric fitness only, we called it *gfit*. A *Source* point matches a *Target* point if their distance is less than a defined threshold. The value *gfit* is the number of matching points in the *Source* cloud divided by its total number of points.

$$gfit = \frac{number-of-matching-points-in-Source}{Total-number-of-points-in-Source} \qquad (8)$$

The RMSE of all matching points is also retrieved as well as the transformation matrix from the *evaluate registration* function. Then, the transformation matrix is applied to point cloud and coordinates of the *Source* molecule to visualize the superimposition of the point cloud and 3D graph, respectively. In this study, PyMOL version 1.3 was used to visualize alignments and generate pictures.

In this study, two other scores are calculated, *cfit* and *hfit*. *cfit* measures the matching of points in each class, whereas *hfit* calculates the same score but without class 1. *hfit* intends to specifically evaluate the matching of polar and aromatic points (class 2, 3 and 4).

$$cfit = \frac{\sum_{i=1}^{k} nb-matching-points-in-Source-for\ class-i}{\sum_{i=1}^{k} number-of-points-in-class-i} \qquad (9)$$

with k the total number of classes (in our study, k = 4; class 1 [H, Cl, Br, I], class 2 [N, O, S, H(NH/OH), F], class 3 [C, P,B] and class 4 [other atoms]).

$$hfit = \frac{\sum_{i=2}^{k} nb-matching-points-in-Source-for\ class-i}{\sum_{i=2}^{k} number-of-points-in-class-i} \qquad (10)$$

Those scores are similar to a Tversky coefficient (18) tuned to evaluate the embedding of one object (here the *Source* molecule) into another one (here the *Target* molecule).

In the following study, the term COLOR *gfit*, *cfit* or *hfit* means that GLOBAL + COLOR methods are applied successively before calculating fitness scores.

**IV - Test cases**

**Figure 2** displays the structures of molecular pairs used to set up and evaluate SENSAAS. Among trivial test cases (**Fig. 2** a), Sorbate(moved) and Imatinib(moved) are 3D structures simply rotated and translated away from the original pose. In such test cases, and if SENSAAS succeeds, the resulting superimposition must be perfectly achieved and the score must equals 1. Substructures pairs like Sorbate/SorbateC, Imatinib/Imatinib-part1, -part2 and –part3 are test cases whose resulting alignment are also easy to validate (**Fig. 2** b). Then, we wanted to align some molecules containing the bioisosteric tetrazole/carboxylate functions (19,20) (**Fig. 2** c). Indeed, we aim at developing a method that performs well in scaffold hopping (*i.e.* replacing one chemical group by another one with a different chemotype) (21). This implies that SENSAAS must not only successfully align molecules of the same size but also must correctly align substructures or bioisosteric fragments, even if they are small. We selected three molecules: Adapalene, which contains a carboxylic acid function, and Irbesartan and Valsartan, which contain a tetrazole ring. Adapalene is more dissimilar to Irbesartan and Valsartan than Irbesartan is to Valsartan. Irbesartan and Valsartan belong to the same therapeutic class of angiotensin II receptor antagonists. The e-Drug3D website http://chemoinfo.ipmc.cnrs.fr/MOLDB/browse.php?query=_718_796_776 (see the Similarities tab) displays alignments and similarity scores calculated by using the program ROCS (22-24).

Resulting scores indicate a ROCS ComboScore of 0.499 and 0.377 between Adapalene and Irbesartan and Valsartan, respectively (scores ranges between 0 and 1). Irbesartan and Valsartan are effectively more similar with a score of 0.701. By aligning these molecules, we aim at evaluating the balance between the global geometry matching and the colored patch matching. Finally, we aligned Tranylcypromine and Milnacipran which are neither therapeutically nor structurally similar but which share a common lipophilic/aromatic substructure (http://chemoinfo.ipmc.cnrs.fr/MOLDB/browse.php?query=_541_288) (**Fig.2** d). Finally, we aligned four structural conformers of the molecule Bictegravir that were generated by using Corina (25) and Omega (OpenEye Software (24,26)) tools for enumerating ring conformers and acyclic rotamers, respectively.

**RESULTS**

Open3D registration methods are generalist for 3D data processing. Thus, registration methods may work with a 3D model of a building or a molecule, regardless of the scale of the object. Therefore, we investigated editable parameters in GLOBAL and COLOR methods and assessed their effects on molecular shape alignment results. **Table 1** displays the list of parameters that can be modified and optimized.

**Optimization of the voxel size**

In our study, points are uniformly distributed on the vdW surface when using the program nsc (11). The distance between two points in the cloud is ~ 0.3. Before performing registration, point clouds are commonly down-sampled to reduce the number of points in the point cloud, and thus, the computation times. The down-sampling is done according to a regular voxel grid and operates in two steps:

- Points are bucketed into voxels;

- Each occupied voxel generates one point by averaging all points inside (*i.e.*, centroid).

As the voxel size is unitless, it appears crucial to set values appropriate for our 3D molecular shapes. **Figure 3** displays the results of the down-sampling on the molecule indoxam when using a voxel size of 0.5, 1.0 or 2.0. **Table 2** shows the average percentage of remaining points for various voxel sizes. As expected, a voxel size $\leq 0.3$ has a limited effect on decreasing the initial number of points since more than 70% of points are kept. On the contrary, a voxel size $\geq 1$ results in pruning more than 90% of initial points. However, if too many points are removed, too much information about geometry and color distribution (color patches) are lost. Thus, we analyzed pairwise alignments for voxel size values ranging from 0.2 to 1.2 with an increment of 0.1. Of note, the same voxel size is applied to both *Target* and *Source* and to both GLOBAL and COLOR methods. In **Figure 4**, three pairs of molecules sorbate/sorbate(moved), Imatinib/Imatinib(moved), Imatinib/Imatinib-part2 are aligned and *gfit* scores are plotted in function of the voxel size. Of note, Sorbate(moved) and Imatinib(moved) structures are the same as Sorbate and Imatinib structures (same conformer), respectively, except that they were rotated and translated away from the original pose. In such test cases, the resulting superimposition must be perfectly achieved by the alignment tool and the score must equals 1. Imatinib-part2 is a substructure of Imatinib (**Fig. 2**). Three runs were carried out. In all runs and whatever the voxel size, the GLOBAL *gfit* score is lower than the COLOR *gfit* score as well as the RMSE (**Fig. 4** a, b and c). This result confirms that the GLOBAL alignment is refined by the COLOR method. Results also show that there is no single value of voxel size for which the best solution could be found for all cases. For example, perfect superimposition can be observed for the sorbate case but not for imatinib when voxel size equals 0.2 and 0.3. The down-sampling when the voxel size equals 0.2 is quite limited (95-96% remaining points) and it seems that a large number of points affect the GLOBAL method result. Indeed, the imatinib point cloud contains ~7800 points whereas the sorbate point cloud contains ~2500 points. In the substructure alignment case (**Fig. 4** c), GLOBAL and COLOR *gfit* scores vary from one test to another but good and similar alignments and scores are obtained for voxel sizes of 0.4, 0.6, 0.7 and 0.8. In this test case, voxel size larger than 0.8 lead to more erratic results. In conclusion, best values of voxel size likely depend on the size, geometry and color distribution as well as differences between the two point clouds. Since we do not know this information in advance and given that the down-sampling is reproducible for a given point cloud and for a given voxel size, we choose to execute the GLOBAL and COLOR methods with each voxel size ranging from 0.2 to 1.2 with an increment of 0.1. This allows initiating the alignment procedure with eleven down-sampled point clouds that retain sufficient information on geometry and color distribution. The best COLOR *gfit* score is selected as the best alignment for the two point clouds.

**Optimization of parameters in GLOBAL**

The *ransac_based_on_feature_matching* function was modified to tune the *RANSACConvergenceCriteria* parameters, *max_iteration* and *max_validation*, which set the maximum iteration before algorithm stops, and maximum times the validation has been run. In RANSAC, most of iterations do not successfully pass the validation step, the most computational step. Compared to the Open3D example, we increased *max_validation* to 1000 but decreased *max_iteration* to 400 000 to keep an acceptable computation time.

**Optimization of parameters in COLOR**

The *colored_icp* registration method was modified to set the *ICPConvergenceCriteria* parameter *max_iteration* to 100 to better optimize the refinement. Values larger than 100 did not improved results in our test cases. A SENSAAS alignment only takes seconds but the current computational cost is still expensive if one wants to perform millions of pairwise alignments.

**Optimization of parameters for evaluation**

The evaluation is performed on the entire dataset of points. The parameter threshold is the maximum correspondence points-pair distance to consider a pairing during evaluation. This parameter is not taken into account in GLOBAL or COLOR. Since the distance between two points in our point clouds is 0.3, we set a threshold value to 0.3. Increasing this threshold would automatically and artificially increase *gfit* scores by pairing more distant points.

**Pairwise alignments**

Molecular pairs from **Figure 2** were aligned by using SENSAAS. Molecular set a) was used to set parameters and successful superimpositions were obtained (**Fig. 4** a and b). Molecular set b) was used to investigate substructure matchings. Three imatinib substructures (only one conformer) were aligned on the target Imatinib. Of note, the pyridine ring in Imatinib-part2 shows a 180° rotation compared to the Imatinib's pyridine conformer. **Figure 5** displays the three superimpositions along with the COLOR *gfit* score. Substructure matching succeeds since each part aligns with its counterpart in Imatinib structure. A small shift is observed for the Imatinib-part1 alignment because of the pyridine conformer difference. Imatinib-part2 is perfectly aligned. Imatinib-part3 superimposition displays a shift in the pairing of the piperazine rings. This is also due to a different conformer of the piperazine ring. Sorbate/SorbaceC is also a trivial test case that SENSAAS successfully aligns.

In this study, we were particularly interested in the bioisosterism between the tetrazole and carboxylate function (molecular set c) in **Figure 2**). Indeed, although these two functional groups appear to be structurally quite different, several studies have demonstrated that they may be interchangeable in a bioactive molecule (19,20). **Figure 6** a) displays the superimposition of these two chemical groups. Aligned point clouds in b) are colored by the standard color element of their closest atoms (nitrogens in blue, oxygens in red, carbons in green and hydrogens in white). In SENSAAS, blue and red points are merged into the same class number 2 since N, H in N-H, O and H in O-H are considered as polar points. Thus, with this coloration, point clouds of these two fragments look alike and SENSAAS properly aligns the three color patches (class 1 in white, class 2 in red, class 3 in green; **Fig. 6** c). Then, we checked that SENSAAS succeeds to align the tetrazole chemical group on the two drugs Irbesartan and Valsartan which contain such a substructure (**Fig. 6** d and e). As well, the pairwise alignment between a carboxylic acid group and the structure of Adapalene leads to the expected superimposition (the same result is obtained when using the carboxylic acid form) (**Fig. 6** f). Then, we investigated whether a tetrazole group can align on the carboxylic acid function of Adapalene or on the carboxylate of Adapalene in its ionized form. **Figure 6** g) and h) display pairwise alignments in which the tetrazole effectively superimposes well with the carboxylate or carboxylic function. An alternate superimposition is sometimes observed in which the methyl group of the tetrazole is aligned on the methoxy group of Adapalene (**Fig. 6** i)). This alignment is scored with as slightly better value of COLOR *gfit*, 0.612 instead of 0.603, but the value of the COLOR *hfit* score, 0.041 instead of 0.584, clearly indicates a mismatch of colored points in this alternate solution.

To complete this study, we carried out pairwise superimpositions between the three drugs to investigate whether bioisosteric groups affect the geometrical alignment. In SENSAAS, the color information is used in a second step to refine the initial alignment from the *Global registration* method. Results show that biososteric groups do not superimpose in any of the two cases (**Fig. 7** a and b) but that the overall geometrical alignment is achieved. COLOR *gfit* values are low but they are consistent with low ROCS ComboScore values displayed in the e-Drug3D database (http://chemoinfo.ipmc.cnrs.fr/MOLDB/browse.php?query=_718_796_776; see Similarities tab). Such values highlight the structural and geometrical dissimilarities of Adapalene with Irbesartan and Valsartan. Moreover, these drugs do not bind the same protein and binding pocket, retinoic acid and angiotensin II receptors, respectively. On the other hand, the pairwise alignment of Valsartan and Irbesartan shows the expected superimposition of the similar phenyl-phenyl-tetrazole substructure (**Fig. 7** c).

An additional pairwise alignment was performed on two drugs, Tranylcypromine and Milnacipran, which do not bind the same protein target but display a common substructure (**Fig. 7** d). Here again, SENSAAS is able to identify and align similar sub-shapes that leads to a good superimposition of the sub-structures of the 3D graph of the molecules even if initial point clouds are significantly different (**Fig. 7** d). Since SENSAAS *gfit* scores are only calculated with the number of pairing points in *Source*, a different *gfit* score is obtained when the opposite calculation is carried out with the *Target* becoming the *Source*. Of note, the opposite calculation leads to the same structural alignment except in cases where several solutions (several local alignments) exist (usually, in such cases, *gfit* scores are low). The smallest molecule of the two will always obtain the highest *gfit* score since, proportionally, more points are paired. We have, on purpose, developed an evaluation mode version of SENSAAS to evaluate the symmetric value.

In the following example, we used SENSAAS to align four conformers of the drug Bictegravir (**Fig. 2** e). Conformers are characterized by structural changes in the oxazepine ring and in two rotamers of the amide linker. Four pairwise alignments with the first conformer as the *Target* is displayed in **Figure 8**. Alignments show a superimposition of conserved substructures, confirming the efficiency of SENSAAS in identifying common sub-shapes in point clouds.

Finally, a preliminary analysis of the distribution of COLOR *gfit* scores was carried out by plotting the values of 500 000 different pairwise alignments between drugs extracted from the e-Drug3D dataset (one conformer; (27))(**Fig. 9**). Drug structures in that dataset are structurally various ranging from small compounds like Histamine to high molecular weight peptides like Triptorelin. Score values range from 0 to 1. Approximately 63% of *gfit* values are in the range ]0.2 – 0.4] and 22% possess a value larger than 0.4, 7% a value larger than 0.5 and only 2% a value larger than 0.6. COLOR *gfit* scores of 1 are obtained for pairwise alignment of structures on themselves. The distribution stringently distinguishes similar from dissimilar molecules. For now, a *gfit* score larger than 0.5 seems to translate into a clear similarity with reproducible results. On the contrary, a *gfit* score lower than 0.5 means a weak molecular similarity and, often, also leads to several (local) solutions when the calculation is repeated. These discriminating scores can be an advantage when one wants to screen millions of structures to select only few similar molecules to the query. A more detailed analysis of *gfit*, *cfit* and *hfit* scores will be performed in the future.

In conclusion, we show that SENSAAS performs well in aligning globally drug-sized molecules as well as in aligning locally substructures, fragments and bioisosteric groups.

**DISCUSSION**

In the present study, we establish that functions *Global registration* and *Colored point cloud registration* from the open-source Open3D librairies can be combined and configured to create an efficient tool for aligning colored point-based surfaces of molecules. In this area of research, Baum *et al.* showed the most accomplished work so far (5,28,29). In their studies, hydrogen atoms are omitted and the solvent excluded surface is partitioned into equally weighted patches represented by a centered point. A point positioning optimization then leads to a point cloud of few hundred points with a neighbored point distance of 2.0 Å (5). Another main difference with our algorithm is that they create distinct point clouds for molecular shape and for each physico-chemical property. In our study, the average neighbored point distance is 0.3 on the vdW surface. This results into a finer resolution of the shape but also into many more points in the cloud. However, nowadays, 3D data processing libraries such as Open3D uses to handle much larger point clouds (representing buildings for example) than point clouds representing the surface of a molecule. In our approach, we also choose to color points in a basic and intuitive way with only four colors representing the physico-chemical property of their closest atoms: polar, apolar, aromatic and other. Such features are often used in 3D pharmacophore designing (30). This way of coloring can easily evolve towards more complexity (more classes, computed physico-chemical properties such as the partial charge…). However, we anticipate that increasing the number of colors will lead to point clouds with many small patches and this will probably make the alignment more difficult to optimize. Further studies are needed to determine these limitations. Our current setting is also insensitive to tautomeric and ionization forms of the molecule as long as few hydrogen atoms are involved. We showed that such changes do not significantly modify the point cloud. This can be another advantage of SENSAAS since 1) such information on molecules is not always known and not easily calculated and, 2) enumerating tautomers considerably increases the size of collections. Then, regarding the molecular similarity score, it would be interesting to combine the three *gfit*, *cfit* and *hfit* scores to discriminate alternate alignments when they exist. For example, we showed that the *hfit* score in the pairwise alignment of Adapalene/Tetrazole helps in selecting the best superimposition (*i.e.* bioisosteric groups alignment) and distinguishing the alternate, geometrical, superimposition.

In conclusion, SENSAAS is able to identify and align similar shapes and sub-shapes that lead to a good superimposition of the structures and sub-structures of the corresponding 3D graphs even in cases where initial point clouds are significantly different. Still, a broader study would be important to assess performance when compared to other tools, especially other 3D shape-based approaches (31,32). Finally, the program, although fast, would also require an optimization of the computational cost for its use in virtual screenings of millions of molecules.

**CONCLUSIONS**

Molecular similarity is a central concept in drug discovery. A wide range of methods have been developed to describe molecules and assess similarity by using representations such as the molecular formula, the chemical graph in 2D or 3D or the molecular shape. In this study, we investigated the use of newly available open-source libraries in 3D data processing to assess point-based surface similarities of molecules. SENSAAS remains close to the concept of the 3D pharmacophore but differs by also taking into account for the geometry of the shape. The method was evaluated and validated against several test cases ranging from pairwise alignments to substructure matching of fragments. It shows potential for molecular similarity evaluation and scaffold hopping. In particular, as it uses open-source libraries and programs, it can be easily retrieved and deployed for other pairwise comparison of shapes such as peptides or proteins.

# FIGURE LEGEND

**Figure 1** The molecule indoxam with or without hydrogens. Hydrogens are colored in white, carbons in green, nitrogen in blue and oxygens in red. Dotted boxes highlight differences between point clouds. a) 3D graph structure of indoxam with hydrogens. b) point cloud generated by using a). c) point cloud generated by using a) in which polar hydrogens are colored with the color of the linked atom. d) 3D graph structure of the ionized indoxam with hydrogens. The carboxylic acid is transformed into a carboxylate function. e) point cloud generated by using d). f) 3D graph structure of indoxam without hydrogens. g) point cloud generated by using f).

**Figure 2** Structures of molecules used as test cases. a) molecular dataset to set parameters. Sorbate(moved) and Imatinib(moved) are 3D structures simply rotated and translated away from the original pose. b) molecular dataset to validate the method on sub-structure test cases. Part1, 2 and 3 are sub-structures of the Imatinib structure. SorbateC structure possesses an additional carbon compared to Sorbate. c) As we are interested in bioisosterism and scaffold hopping, we selected some examples in which the tetrazole/carboxylate bioisosterism can be challenged. Adapalene, Irbesartan, and Valsartan possess either of the chemical functions. d) Tranylcypromine and Milnacipran also share a common substructure although they do not possess similar bioactivities. e) The drug Bictegravir is used as a test case to align its conformers.

**Figure 3** Effect of the voxel size on the number of points in the down-sampled point cloud. The original point cloud was calculated using the molecule indoxam.

**Figure 4** GLOBAL and COLOR *gfit* scores in function of voxel sizes ranging from 0.2 to 1.2. Three runs are plotted (blue, orange and mauve lines). The GLOBAL initial alignment and the best final alignment are displayed along with their COLOR *gfit*, *cfit*, *hfit* scores and RMSE value. *Target* and *Source* 3D structures are colored in green and cyan, respectively and the *Source* structure after GLOBAL is colored in magenta. a) Imatinib/Imatinib(moved) pair. b) Sorbate/Sorbate(moved) pair. c) Imatinib/Imatinib-part2 pair.

**Figure 5** Pairwise alignments of the four test cases used to validate substructure matchings. a) Imatinib/Imatinib-part1. b) Imatinib/Imatinib-part2. c) Imatinib/Imatinib-part3. d) Sorbate/SorbateC. *Target* and *Source* 3D structures are colored in green and cyan, respectively.

**Figure 6** Pairwise alignments to evaluate the bioisosterim of tetrazole/carboxylate chemical groups. *Target* and *Source* 3D structures are colored in green and cyan, respectively. a) pairwise alignment of the tetrazole function with a carboxylate function. b) Aligned point clouds are colored by the standard color element of their closest atoms (nitrogens in blue, oxygens in red, carbons in green and hydrogens in white). c) Aligned point clouds are colored by classes in SENSAAS (class 1 in white, class 2 in red and class 3 in green). d) pairwise alignment of Irbesartan with the tetrazole group. e) pairwise alignment of Valsartan with the tetrazole group.

f) pairwise alignment of Adapalene with a carboxylic acid group. g) pairwise alignment of Adapalene in a carboxylate form with the tetrazole group. h) pairwise alignment of Adapalene in a carboxylic form with the tetrazole group. i) Alternate pairwise alignment of Adapalene with the tetrazole group.

**Figure 7** Pairwise alignments of structurally dissimilar molecules. *Target* and *Source* 3D structures are colored in green and cyan, respectively. a) Adapalene/Irbesartan. b) Adapalen/Valsartan. c) Valsartan/Irbesartan. d) Tranylcypromine/Milnacipran. Point clouds of Tranylcypromine and Milnacipran are displayed. Point clouds are colored by the standard color element of their closest atoms (nitrogens in blue, oxygens in red, carbons in green and hydrogens in white).

**Figure 8** Pairwise alignments of four conformers of the molecule Bictegravir. *Target* and *Source* 3D structures are colored in green and cyan, respectively.

**Figure 9** Distribution of the COLOR *gfit* scores for 500000 different pairwise alignments of drugs from the e-Drug3D dataset.

# TABLES

**Table 1.** GLOBAL, COLOR and *evaluation* registration parameters. Highlighted cells indicate which parameters were modified in our study when compared to values in Open3D examples.

| Method | Parameter name | Current value(s) | Open3D example (default value) |
|---|---|---|---|
| | vdW surface calculated by using nsc | | the Redwood dataset [Choi2015] |
| **GLOBAL (generates a rough alignment as initialization)** | | | |
| voxel_down_sample() | voxel size (vs) | [0.1, 0.2, 0.3, 0.4, 0.5, 0.6, 0.7, 0.8, 0.9, 1.0, 1.1, 1.2] | 0.05 |
| estimate_normals() | radius_normal | vs * 2 | vs * 2 |
| estimate_normals() | max_nn | 30 | 30 |
| compute_fpfh_feature() | radius_feature | vs * 5 | vs * 5 |
| compute_fpfh_feature() | max_nn | 100 | 100 |
| registration_ransac_based_on_feature_matching() | distance_threshold | vs * 1.5 | vs * 1.5 |
| registration_ransac_based_on_feature_matching() | TransformationEstimationPointToPoint() | False | False |
| registration_ransac_based_on_feature_matching() | ransac_n | 4 | 4 |
| registration_ransac_based_on_feature_matching() | checkers : CorrespondenceCheckerBasedOnEdgeLength | 0.9 | 0.9 |
| registration_ransac_based_on_feature_matching() | checkers : CorrespondenceCheckerBasedOnDistance | distance_threshold | distance_threshold |
| registration_ransac_based_on_feature_matching() | RANSACConvergenceCriteria(max_iteration, max_validation) | 400000, 1000 | 4000000, 500 |
| **COLOR (refine initial alignment by using color)** | | | |
| estimate_normals() | vs | the value used in **Global** | [ 0.04, 0.02, 0.01 ] |
| estimate_normals() | radius size | vs * 2 | vs * 2 |
| estimate_normals() | max nn | 30 | 30 |
| registration_colored_icp() | relative fitness | 1.00E-06 | 1.00E-06 |
| registration_colored_icp() | relative rmse | 1.00E-06 | 1.00E-06 |
| registration_colored_icp() | max iter | 100 | [ 50, 30, 14 ] |
| **Evaluation** | | | |
| | threshold | 0.3 | |

**Table 2.** Percentage of remaining points after down-sampling the molecular shapes.

| Voxel size | Percentage of remaining points |
|---|---|
| 0.1 | 99-100 |
| 0.2 | 95-96 |
| 0.3 | 67-70 |
| 0.4 | 43-46 |
| 0.5 | 29-31 |
| 0.6 | 21-22 |
| 0.7 | 16-17 |
| 0.8 | 12-13 |
| 0.9 | 10 |
| 1.0 | 8 |
| 1.1 | 7-8 |
| 1.2 | 5-6 |
| 3 | 0.9-1.1 |
| 7 | 0.3-0.6 |

# AUTHOR INFORMATION

## Corresponding Author

* To whom correspondence should be addressed, E-mail: douguet@ipmc.cnrs.fr and fpayan@i3s.unice.fr.

## Author Contributions

The manuscript was written through contributions of all authors. All authors have given approval to the final version of the manuscript.

## Funding Sources

This work was supported by the Centre National de la Recherche Scientifique (CNRS) and the Institut National de la Santé et de la Recherche Médicale (Inserm). This work was also supported by the French government through the French National Research Agency (ANR) under the project 'Investissements d'Avenir' UCA$^{JEDI}$ with the reference number ANR-15-IDEX-01.

## ACKNOWLEDGMENT

The authors would like to thank Dr. Antonini, M. (Université Côte d'Azur, CNRS, I3S) for fruitful discussions on the project.

## ABBREVIATIONS

SENSAAS, SENsitive Surface As A Shape; GLOBAL, Global registration; COLOR, Colored point cloud registration.

**Figure 1**

a) 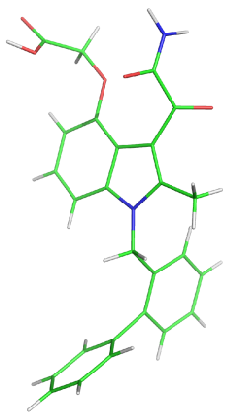

d) 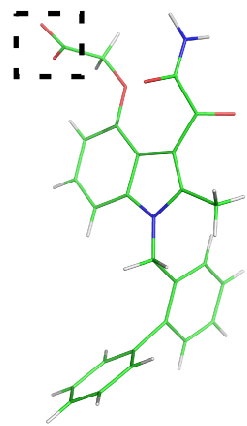

f) 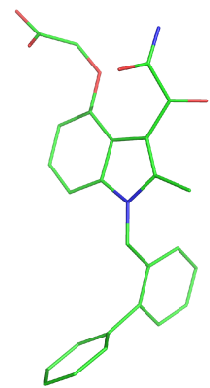

b) 6490 points 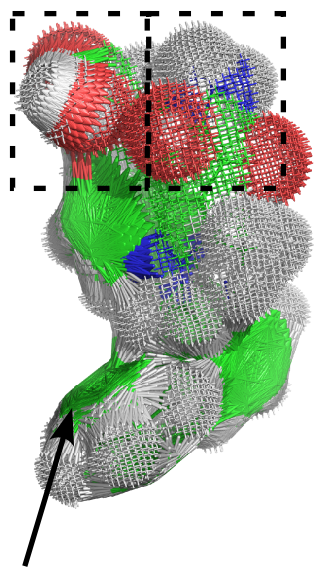

c) 6490 points 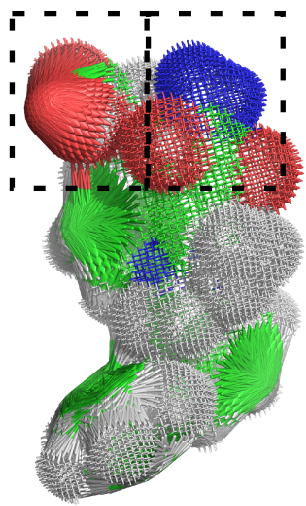

e) 6382 points 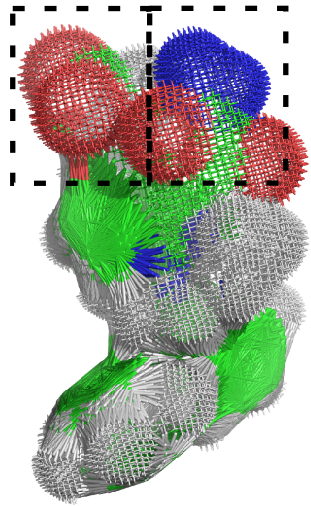

g) 4412 points 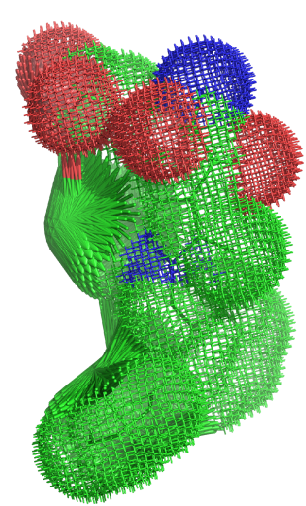

Aromatic green patch

**Figure 2**

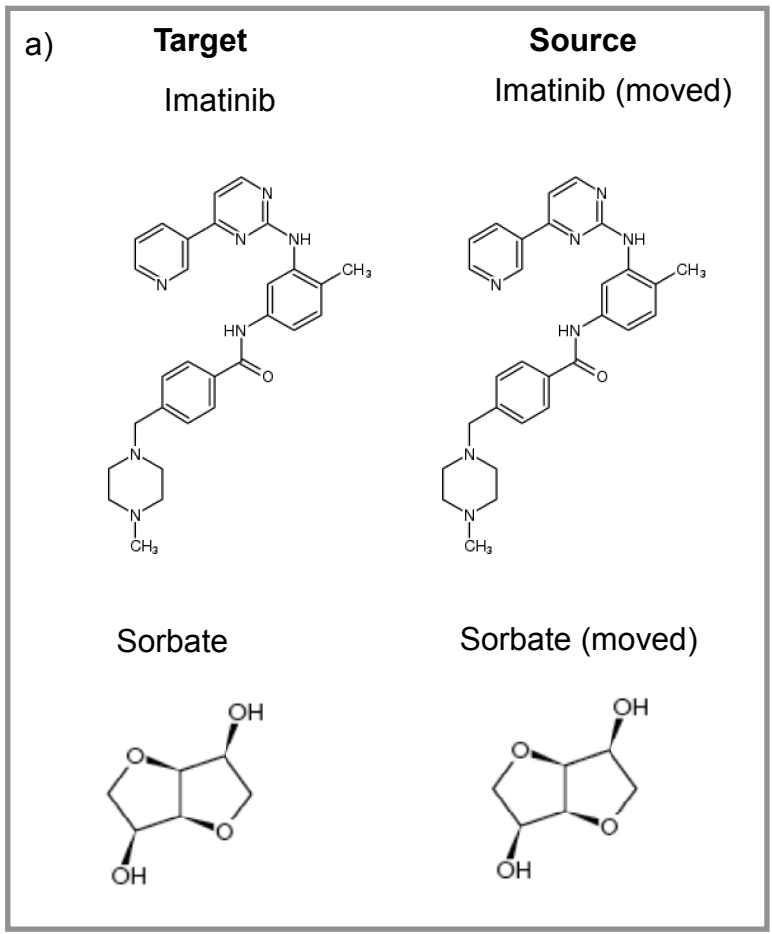
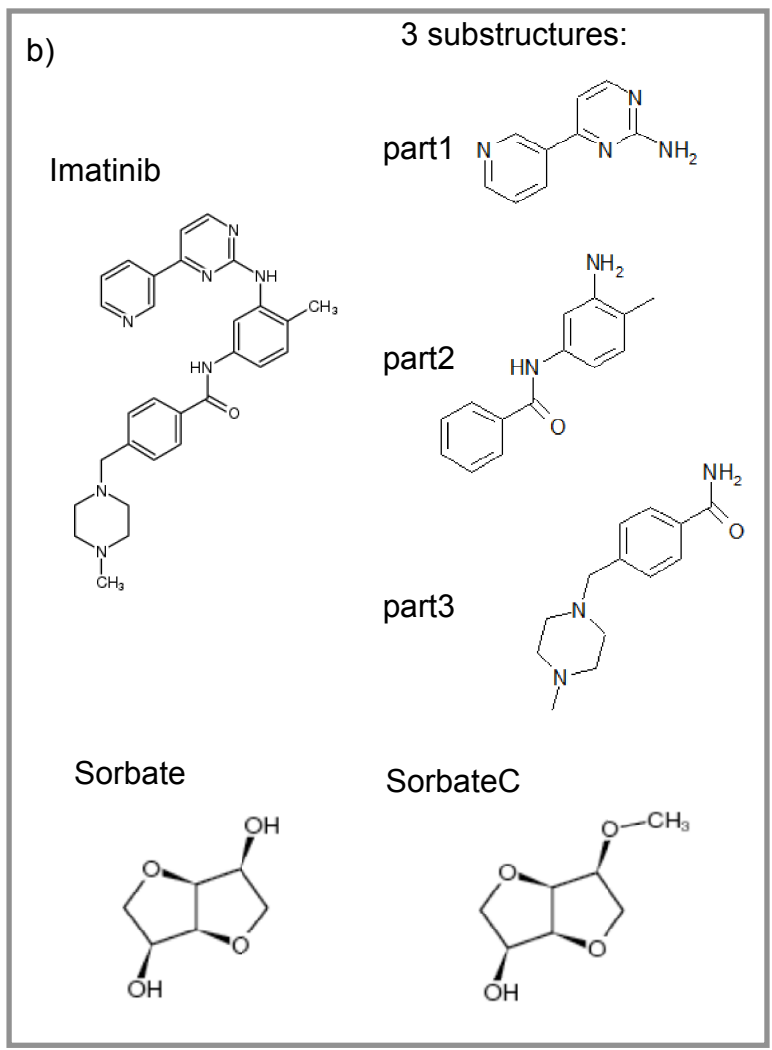
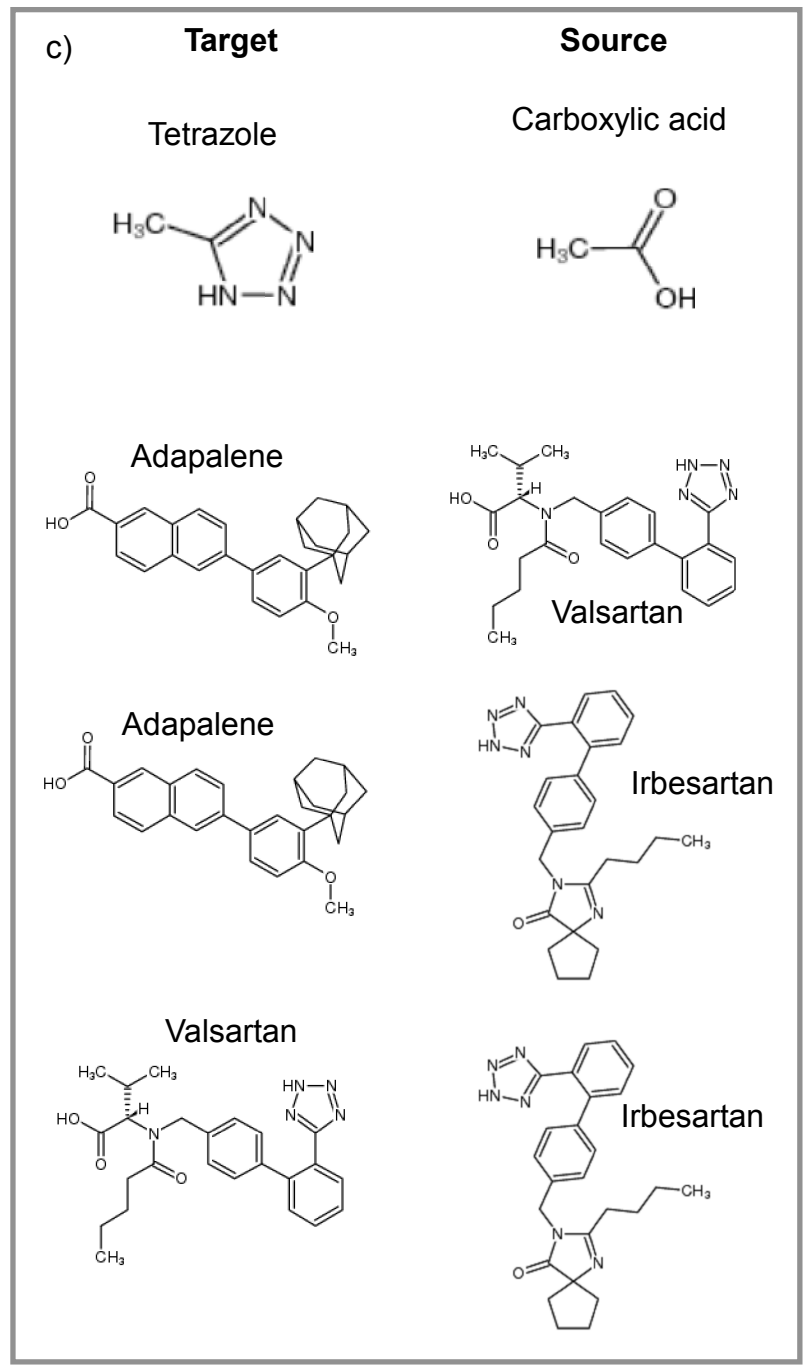
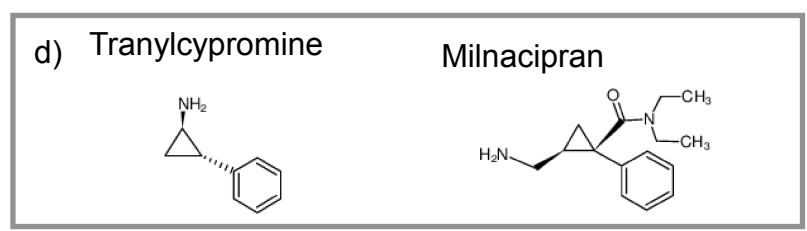
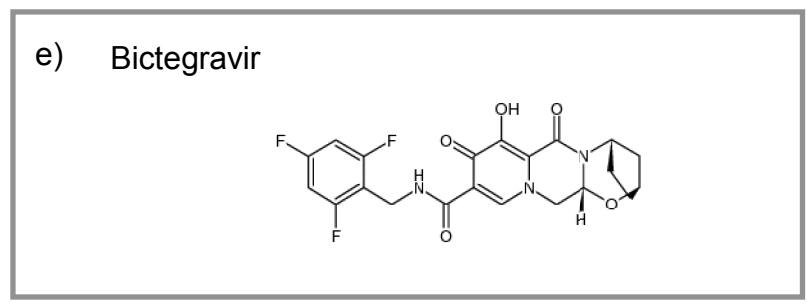

**Figure 3**

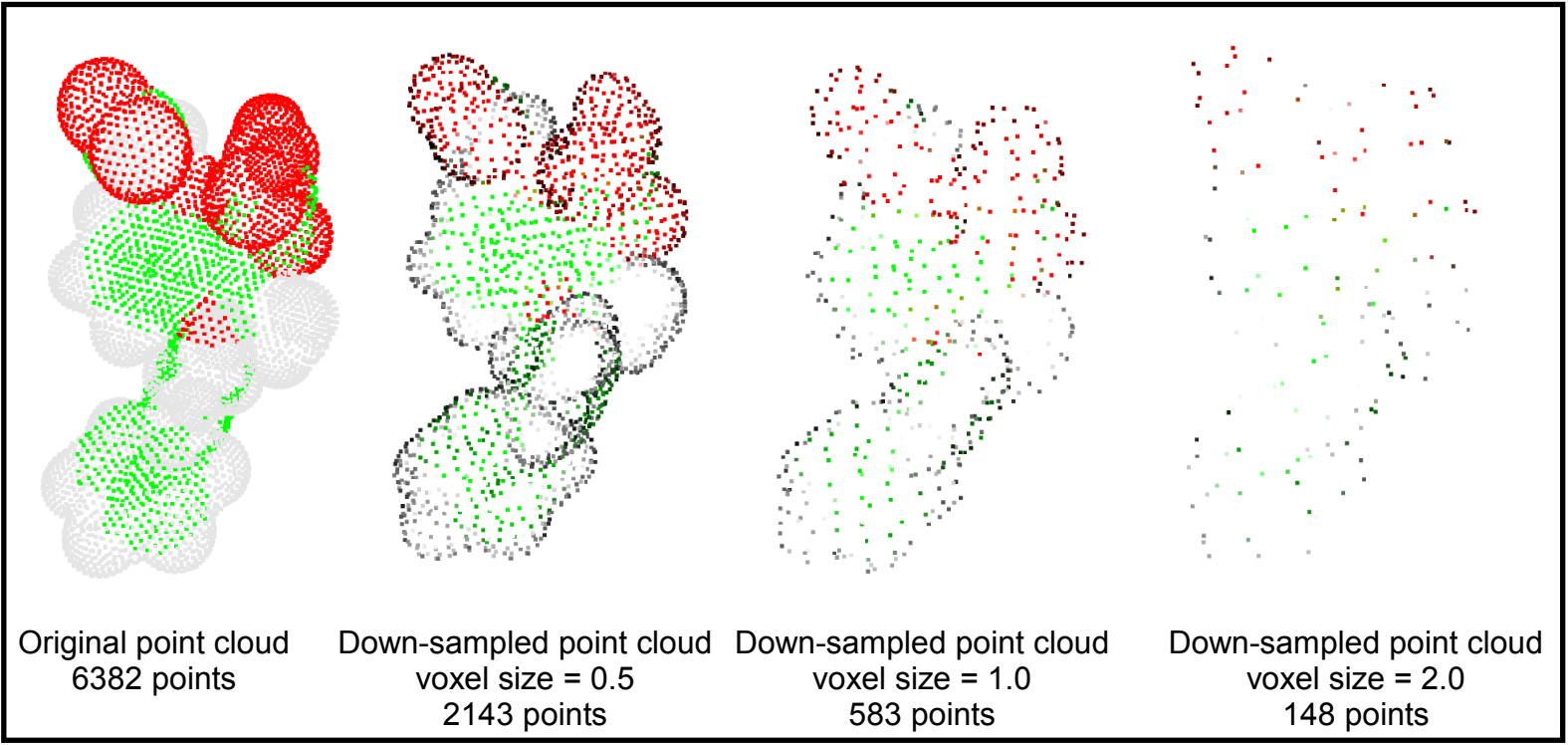

**Figure 4**

### a) Alignment of Imatinib(moved) on Imatinib

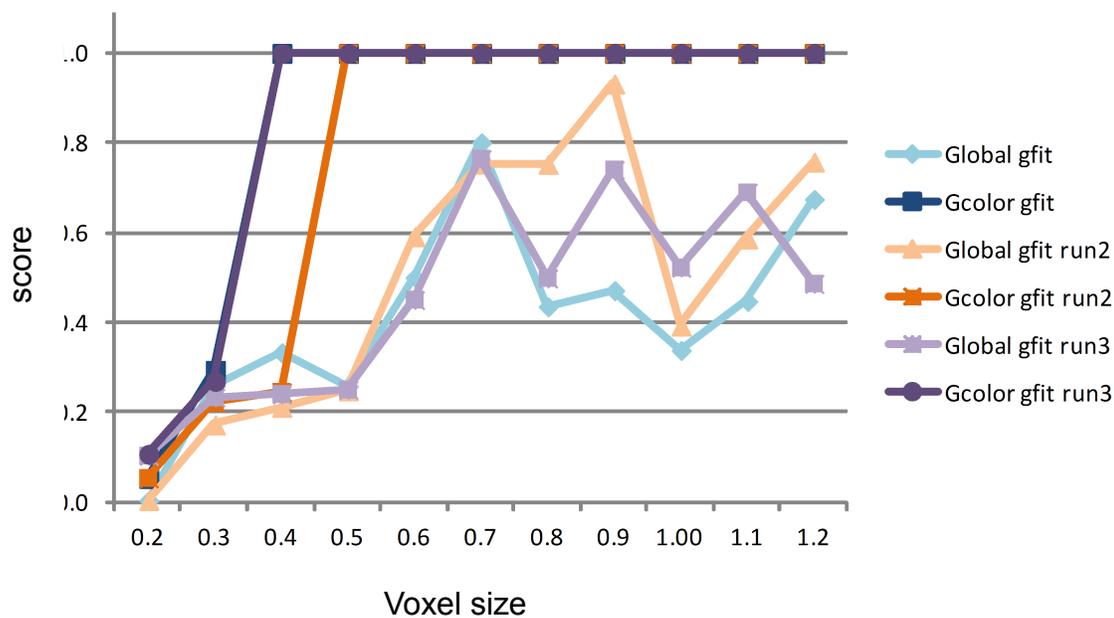

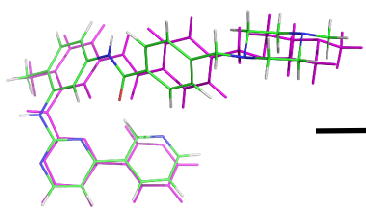

**voxel size = 0.4**

Global alignment result:

GLOBAL gfit = 0.254
RMSE = 0.190

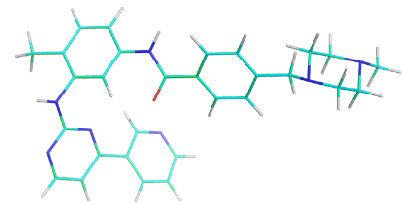

Final alignment:

**COLOR gfit = 1.0**
cfit = 0.998
hfit = 0.996
RMSE = 0.105

### b) Alignment of Sorbate(moved) on Sorbate

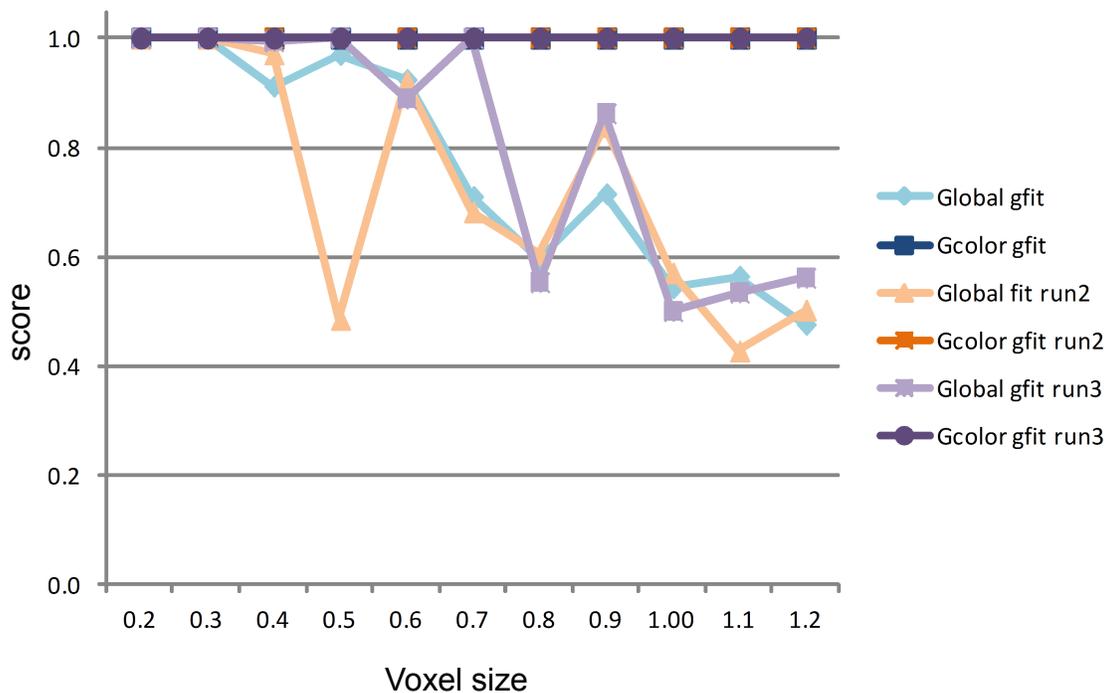

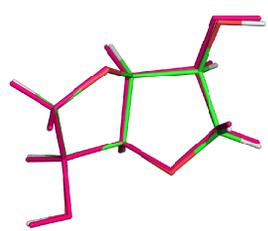

**voxel size = 0.2**

Global alignment result:

GLOBAL gfit = 1.0
RMSE = 0.130

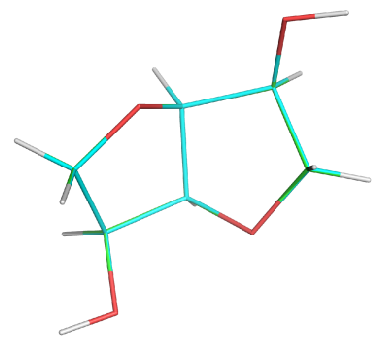

Final alignment:

**COLOR gfit = 1.0**
cfit = 0.999
hfit = 0.999
RMSE = 0.103

**c) Alignment of Imatinib-part2 on Imatinib**

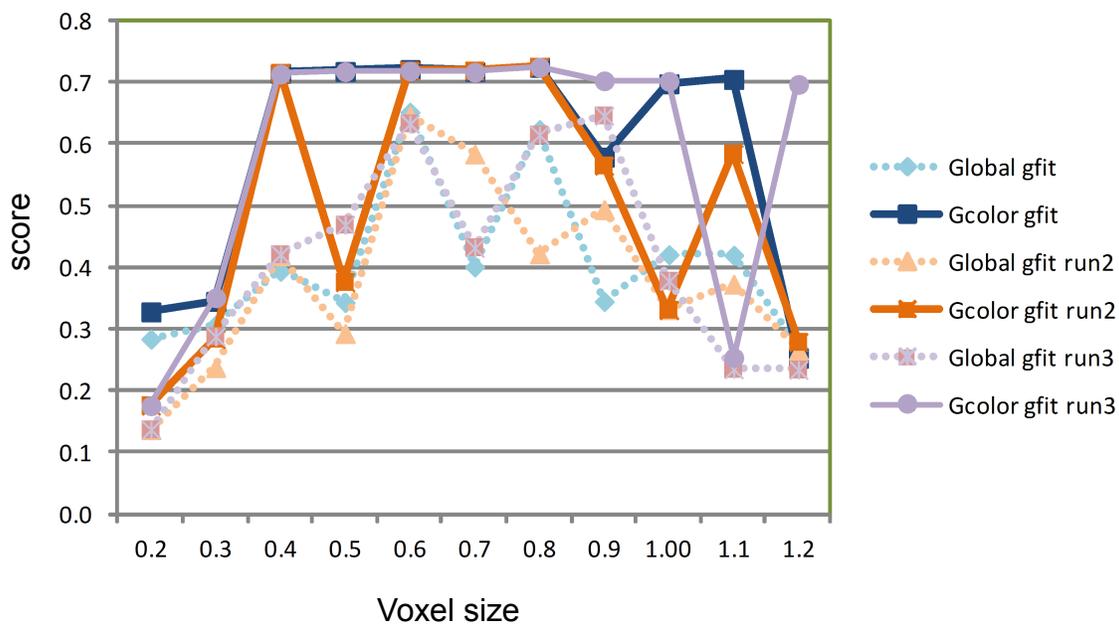

**voxel size = 0.8**

Global alignment result:

GLOBAL gfit = 0.625
RMSE = 0.190

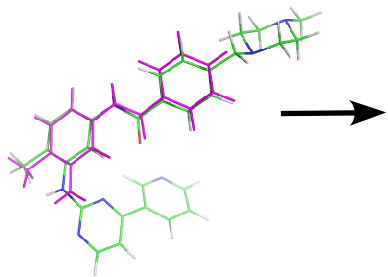

Final alignment:

**COLOR gfit = 0.726**
cfit = 0.665
hfit = 0.638
RMSE = 0.184

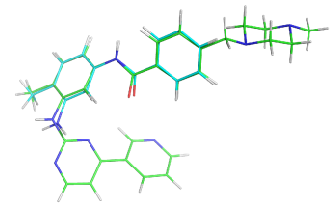

**Figure 5**

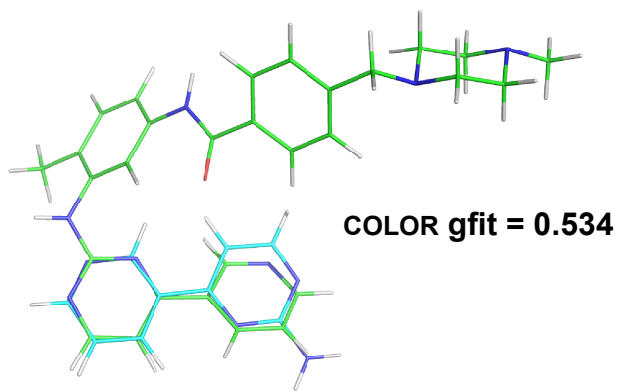
a) Imatinib (Target) / Imatinib-part1 (Source)

COLOR gfit = 0.534

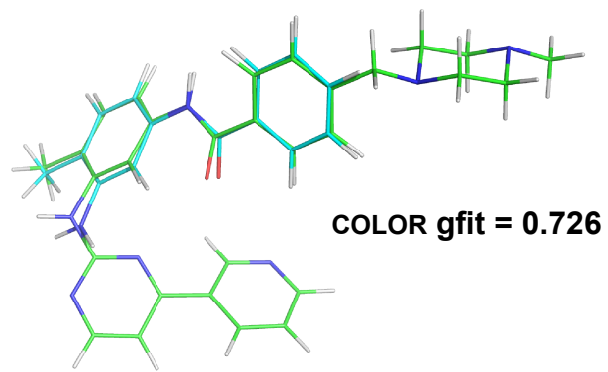
b) Imatinib (Target) / Imatinib-part2 (Source)

COLOR gfit = 0.726

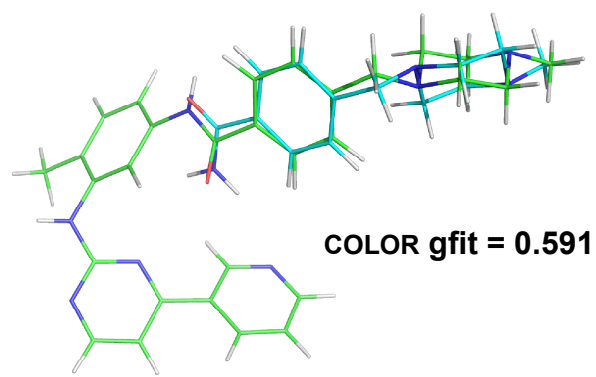
c) Imatinib (Target) / Imatinib-part3 (Source)

COLOR gfit = 0.591

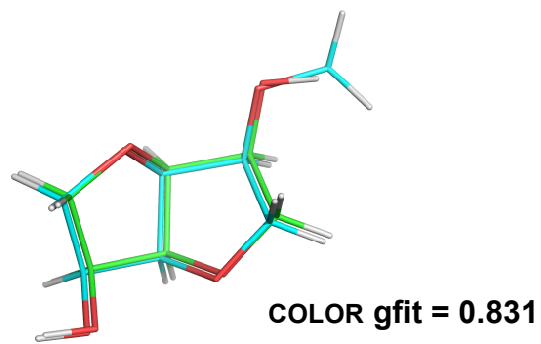
d) Sorbate (Target) / Sorbate-C (Source)

COLOR gfit = 0.831

# Figure 6

## Tetrazole (Target) / Carboxylic acid (Source)

a) 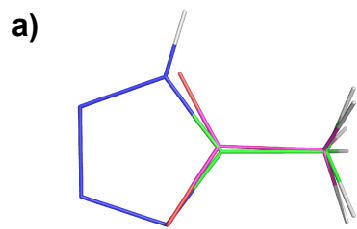

COLOR gfit = 0.836

b) 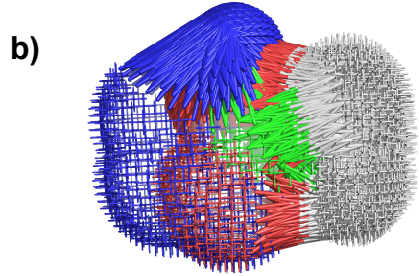

c) 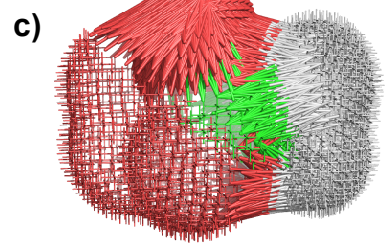

d) Irbesartan (Target) / Tetrazole (Source)

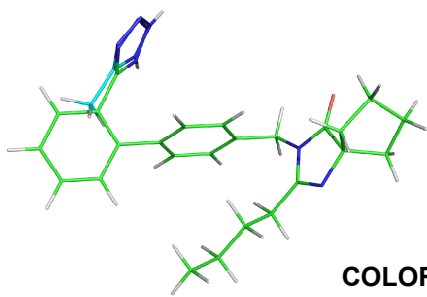

COLOR gfit = 0.672

e) Valsartan (Target) / Tetrazole (Source)

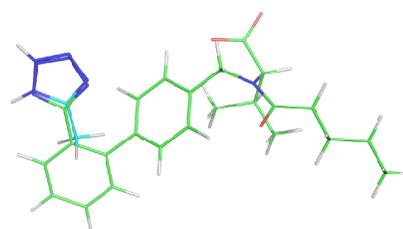

COLOR gfit = 0.690

f) Adapalene (Target) / Carboxylic acid (Source)

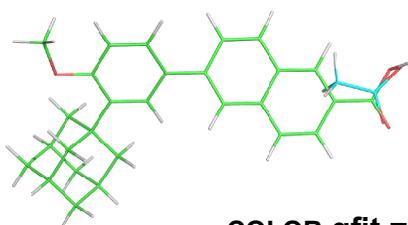

COLOR gfit = 0.686

g) Adapalene - carboxylate form (Target) / Tetrazole (Source)

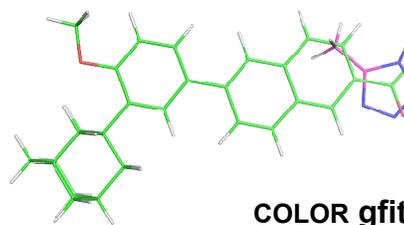

COLOR gfit = 0.606

h) Adapalene (Target) / Tetrazole (Source)

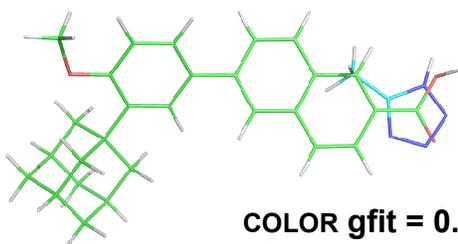

COLOR gfit = 0.603
cfit = 0.431
hfit = 0.584

i) Adapalene (Target) / Tetrazole (Source) Alternate pose

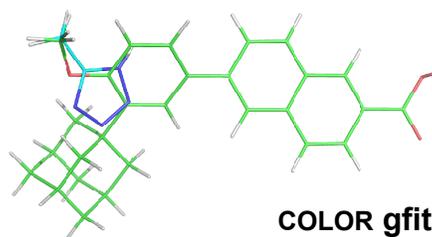

COLOR gfit = 0.612
cfit = 0.427
hfit = 0.041

**Figure 7**

**a) Adapalene (Target) / Irbesartan (Source)**

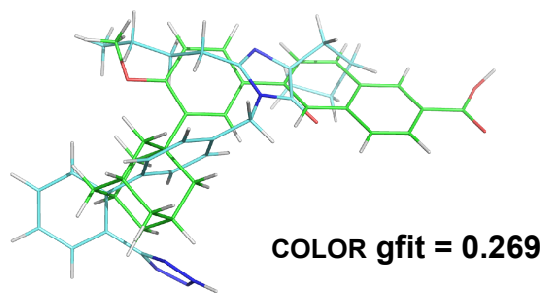

**COLOR gfit = 0.269**

**b) Adapalene (Target) / Valsartan (Source)**

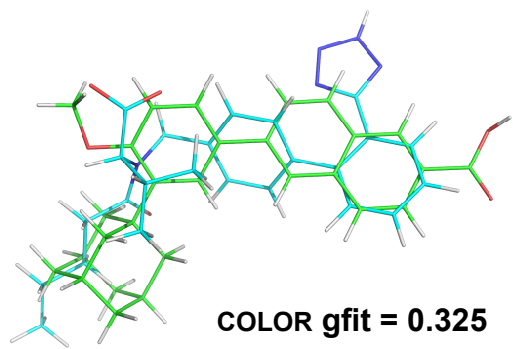

**COLOR gfit = 0.325**

**c) Valsartan (Target) / Irbesartan (Source)**

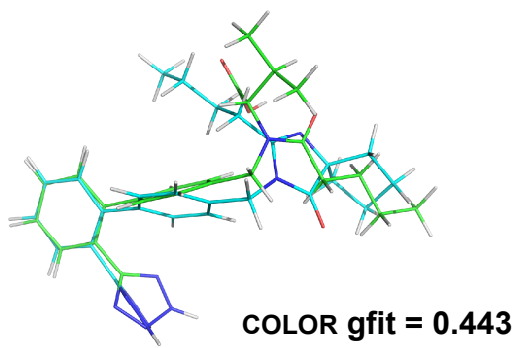

**COLOR gfit = 0.443**

**d) Tranylcypromine (Target) / Milnacipran (Source) - COLOR gfit = 0.384**
   **Milnacipran (Target) / Tranylcypromine (Source) - COLOR gfit = 0.685**

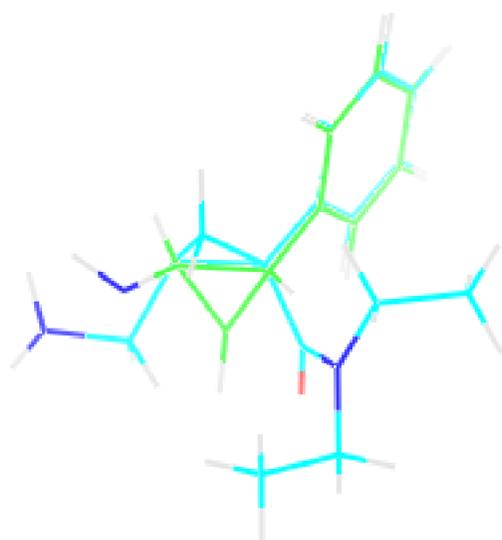

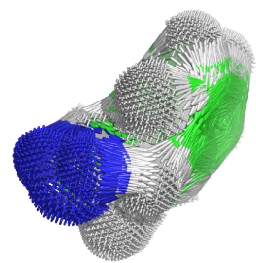

Tranylcypromine point cloud

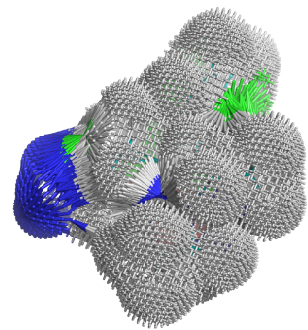

Milnacipran point cloud

**Figure 8**

## Bictegravir (4 conformers)

Conformer 1 / Conformer 1
COLOR gfit = 1

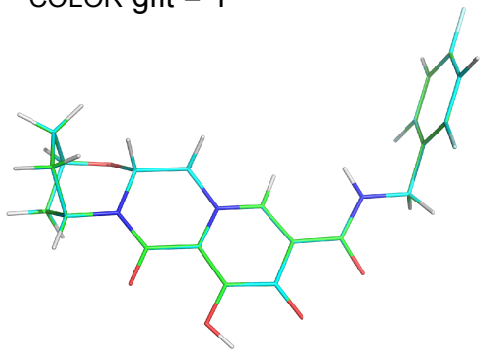

Conformer 1 / Conformer 2
COLOR gfit = 0.660

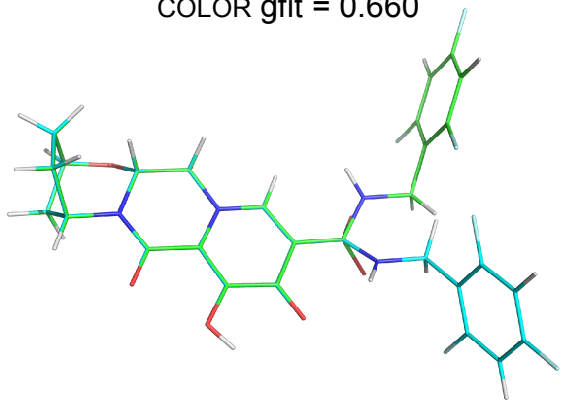

Conformer 3
COLOR gfit = 0.763

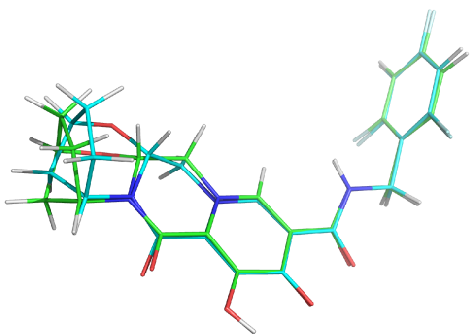

Conformer 4
COLOR gfit = 0.472

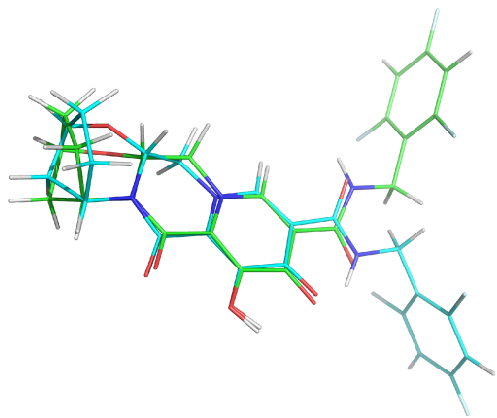

**Figure 9**

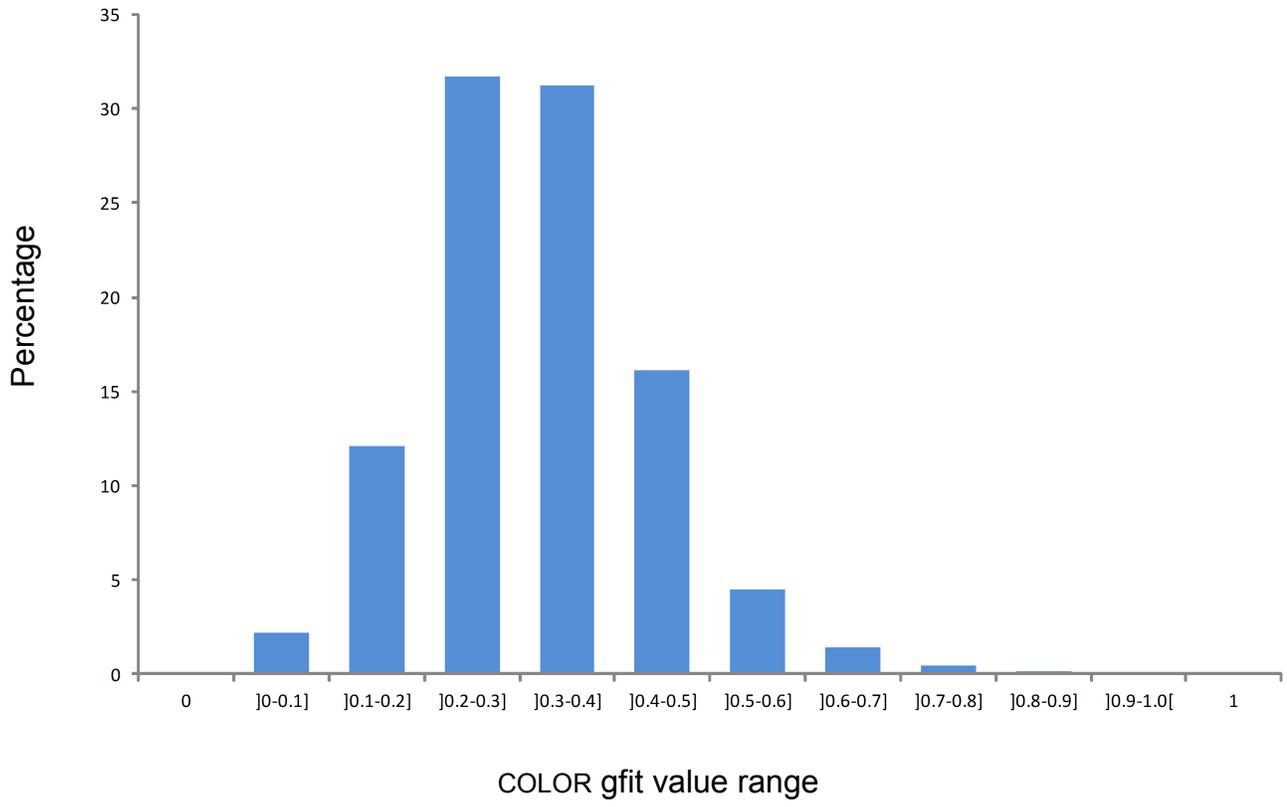